\documentstyle[12pt]{article}
\input epsf
\title{Electrodynamical model of quasi-efficient financial market}
\author{ Kirill N.Ilinski\thanks{E-mail: kni@th.ph.bham.ac.uk}\ 
and\ Alexander S.  Stepanenko\thanks{E-mail: ass@th.ph.bham.ac.uk} \\ [0.2cm]
{\small\it School of Physics and Space Research, University of
Birmingham,} \\ {\small\it Edgbaston B15 2TT, Birmingham, United
Kingdom} }

\begin{document} \maketitle \vskip -9.5cm \vskip 9.5cm

\begin{abstract}
The modelling of financial markets presents a problem which is both
theoretically challenging and practically important.  The theoretical
aspects concern the issue of market efficiency which may even have
political implications~\cite{Cuthbertson}, whilst the practical side of
the problem has clear relevance to portfolio management~\cite{Elton} and
derivative pricing~\cite{Hull}.  Up till now all market models contain
``smart money" traders and ``noise" traders whose joint activity
constitutes the market~\cite{DeLong,Bak}.  On a short time scale this
traditional separation does not seem to be realistic, and is hardly
acceptable since all high-frequency market participants are professional
traders and cannot be separated into ``smart" and ``noisy".  In this
paper we present a ``microscopic" model with homogenuous quasi-rational
behaviour of traders, aiming to describe short time market behaviour.
To construct the model we use an analogy between ``screening" in quantum
electrodynamics and an equilibration process in a market with temporal
mispricing~\cite{Ilinski,Dunbar}.  As a result, we obtain the
time-dependent distribution function of the returns which is in
quantitative agreement with real market data and obeys the anomalous
scaling relations recently reported for both high-frequency exchange
rates~\cite{Breymann}, S\&P500~\cite{Stanley} and other stock market
indices~\cite{Bouchaud,Australia}.  
\end{abstract}

\newpage 

When a mispricing appears in a market, market
speculators and arbitrageurs rectify the mistake by obtaining a profit
from it.  In the case of profitable fluctuations they move into 
profitable assets, leaving comparably less profitable ones.  This affects
prices in such a way that all assets of similar risk become equally
attractive, i.e.  the speculators restore the equilibrium.  If this
process occurs infinitely repidly, then the market corrects the mispricing
instantly and current prices fully reflect all relevant information.  In
this case one sais that the market is efficient.  However, clearly it
is an idealization and does not hold for small enough
times~\cite{Sofianos}. Here, we propose a ``microscopic" model to describe the
money flows, the equilibration and the corresponding statistical
dynamics of prices.

The general picture, sketched above, of the restoration of equilibrium in
financial markets resembles screening in electrodynamics.
Indeed, in the case of electrodynamics, negative charges move into
the region of the positive electric field, positive charges get out of the
region and thus screen the field.  Comparing this with the financial
market we can say that a local virtual arbitrage opportunity with a
positive excess return plays a role of the positive electric field,
speculators in the long position behave as negative charges, whilst the
speculators in the short position behave as positive ones. 
Movements of positive  and negative charges screen out a profitable
fluctuation and restore the equilibrium so that there is no
arbitrage opportunity any more, i.e. the speculators have eliminated the
arbitrage opportunity.

The analogy is apparently superficial, but it is not.  It was shown
in~\cite{Ilinski} that the analogy emerges naturally in the framework of
the Gauge Theory of Arbitrage (GTA).  The theory treats a calculation of
net present values and asset buying and selling as a parallel transport
of money in some curved space, and interpret the interest rate, exchange
rates and prices of asset as proper connection components.  This
structure is exactly equivalent to the geometrical structure underlying
the electrodynamics where the components of the vector-potential are
connection components responsible for the parallel transport of the
charges.  The components of the corresponding curvature tensors are the
electromagnetic field in the case of electrodynamics and the excess rate
of return in case of GTA.  The presence of uncertainty is equivalent to the
introduction of noise in the electrodynamics, i.e.  quantization of
the theory.  It allows one to map the theory of the capital market onto the
theory of quantized gauge field interacting with matter (money flow) fields.
The gauge transformations of the matter field correspond to a change of
the par value of the asset units which effect is eliminated by a gauge tuning of
the prices and rates.  Free quantum gauge field dynamics (in the absence of
money flows) is described by a geometrical random walk for the assets prices with
the log-normal probability distribution. In general case the
consideration maps the capital market onto Quantum Electrodynamics where
the price walks are affected by money flows.

To drop technicalities and put it in simple terms, we consider a composite
system of price and money flows.  In this model ''money" represents
high frequency traders with a short 
characteristic trading time (investment horizon) $\Delta $ (for
the case of S\&P500 below we use 0.5 min as the smallest horizon).
The participants trade with each other and investors with longer
time horizons.  This system is characterized by the joint probability
distribution of money allocation and price.  If we neglect the
money, the price obeys the geometrical random walk which is due to
incoming information and longer time horizons traders.  The 
trader's behaviour on time step $\Delta$ at price $S$ is described by 
the decision matrix of non-normalized transition probabilities~\cite{Ilinski}:  
\begin{equation}
\pi (\Delta) = \left( \begin{array}{cc} 1 & S^{\beta (\Delta )} \\ 
S^{-\beta (\Delta )} & 1
\end{array} \right) 
\label{P} 
\end{equation} where the upper row
corresponds to a transition to cash from cash and shares and lower row
gives corresponding probabilities for a transition to shares.  The
parameter $\beta$ is a fitting parameter playing the role of the
effective temperature. At this stage different traders are independent of each
other. We introduce an interaction by making hopping elements depending 
additionally on change in traders configuration. This interaction models the
``herd" behaviour for large changes and mean-reversion anticipation for small
changes. Each trader possesses lot of shares or the equivalent cash amount. The
formulation of the model is completed by saying that the transition
probability for the market is a product of the geometrical random walk
weight for price and the matrices (\ref{P}) for each participant.

The matrix $\pi (\Delta )$ has exactly the same form as the hopping matrix for
charged particles in Quantum Electrodynamics.  This form can
also be derived from the assumption that traders want to maximize their
profit~\cite{Ilinski}.  This seems to be realistic for the small times
that we are interested in.

It can be shown that a model with just one time horizon cannot
correctly describe effects which have characteristic times of more than 5
minutes.  To improve the model we have to include traders with other
characteristic time horizons.  It is known that there are conventional
intra-day time horizons like 1, 10, 30, 60, 480 mins which, however,
have certain measure of idealization.  We use a continuous set of time
horizons between 30 seconds and 480 minutes to describe the spread and
uncertainty in the time horizons definition.  It means that the model
contains a set of money flows defined by the matrices (\ref{P}) with the
corresponding parameters $\Delta$ and $\beta$. The suggested model
thereby consists of the Fractional Market Hypothesis (FMH)~\cite{Peters}
(which states that a stable market consists of traders with different
time horizons but with identical dynamics) and the ``microscopical"
electrodynamical model for the dynamics.  Hence the FMH substitutes in our
approach the information cascade suggested recently to explain the scaling
properties of the \$/DM exchange rate~\cite{Breymann}.

The use of the FMH is not the only feature of this model which differs from the
ones proposed earlier. The other feature is the homogeneity of the traders
set.  In earlier models traders have always been divided into ``smart" 
(who trade rationally) and ``noisy" (who follow a fad)~\cite{DeLong,Bak}. We
believe that for the consideration of short times trades this
differentiation is not appropriate.  Indeed, all high-frequency market
participants are professional traders with years of experience.
Unsuccessful traders quickly leave the market and do not affect the
dynamics.  At the same time, each of the traders has their own view on the
market and their own anticipations.  That is why their particular decision
can be only modeled in a probabilistic way.  In this sense the traders
are not strictly rational but ''quasi-rational" and the corresponding
market where the {\it quasi}-rational investors deal, can be called a
quasi-effective market.

Let us turn to the results.  First of all, the constructed model allows
us to explain quantitatively the observed high-frequency return data.
In Ref~\cite{Stanley} Figs.1,2 show the form of the distribution
function for changes in the S\&P500 market index, which is a price of the
portfolio consisting of the main 500 stocks traded on the New York
Stock Exchange.  The changes in price have been normalized by the
standard deviation.  In the approximation that the changes are much
smaller than the index itself, which is obeyed with very high accuracy,
the distribution function of the normalized changes can be considered as
the distribution function of the return on the portfolio, normalized by
the standard deviation of the return.  The return on the portfolio
during the period $\Delta$ is defined as $r(\Delta) = (S(t+\Delta)
-S(t))/S(t)$. In
Ref~\cite{Stanley} it was also shown that the distribution function obeys the
scaling property and that this property is reflected in the dependence on time
of the probability to return to the origin. It was demonstrated that
for a time period between 1 min and 1000 min (two trading days) the
probability decrease as $t^{-\alpha}$ with the exponent $\alpha=0.712\pm
0.25$ (see Fig.1).  Similar results have been obtained in
Ref.~\cite{Breymann} for the high-frequency return for the \$/DM exchange rate
with slightly different values of the exponent. We choose parameters of 
our model to get the correct scaling behaviour and define $\beta (\Delta)$ as
$\beta (\Delta)=30/\Delta^{0.71}$. We also take the number of traded lots 
infinite. Fig.1 clearly
demonstrates the correct scaling property of the model constructed above
and gives the same scaling exponent $\alpha=0.71$ which is technically
due to the scaling form of $\beta (\Delta )$. Now we can plot also the 
probability distribution function of returns for S\&P500 as depicted on Fig.2.
The same analysis leads to similar results for the \$/DM exchange
rate~\cite{Breymann} with slightly different values of the parameters. It
is easy to see that the theoretical and observed distribution functions
coincide exactly with the observed data accuracy.  

The shape of the distribution function does not characterize it
completely.  Indeed, a similar form can be obtained using the GARCH/ARCH
models~\cite{Engle,Bollerslev}, which still are phenomenological rather
than microscopic.  However, those models cannot explain the scaling
properties obeyed by the real data~\cite{Breymann,Stanley}.  
In the proposed model both the shape of the distribution function and
scaling properties are presented. 
It is interesting to add
that the swings of the real data on Fig.1 can be interpreted as a sign of
the inhomoginuity of the distribution of traders across time horizons,
with a larger number of traders on 15 mins, 60 mins and a day
investment horizons.

ACKNOWLEDGEMENTS.  We thank R.Mantegna for sending us experimental data used
in Figs 1,2.

\newpage

\begin{center} {Figure caption} \end{center}

\vspace{2cm}

FIG.1 Theoretical (solid line) and experimental (squares) probability of
return to the origin (to get zero return) $P(0)$ as a function of time.
The slope of the best-fit straight line is -0.712$\pm$0.025~\cite{Stanley}.
The theoretical curve converges to the Brownian value $0.5$ as time tends to
one month.

\vspace{2cm}

FIG.2 Comparison of the $\Delta =1$ min theoretical (solid line) and
observed~\cite{Stanley} (squares) probability distribution of the return
$P(r)$.  The dashed line (long dashes) shows the gaussian distribution with the
standard deviation $\sigma$ equal to the experimental value 0.0508.
Values of the return are normalized to $\sigma$.  The dashed line (short dashes) is the
best fitted symmetrical Levy stable distribution~\cite{Stanley}.

\begin{figure}
\centerline{\epsfxsize=16cm \epsfbox{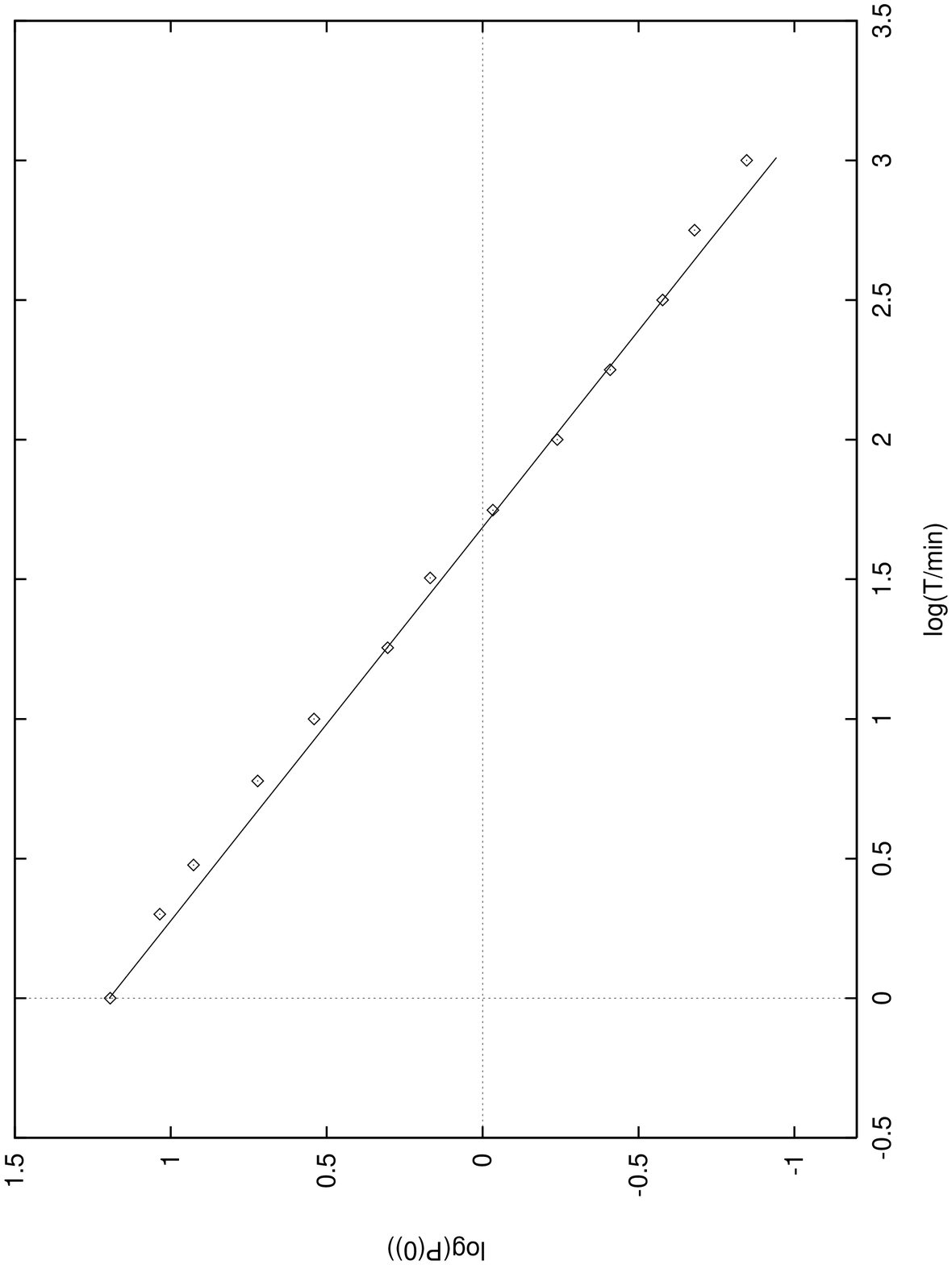}}
\end{figure}

\begin{figure}
\centerline{\epsfxsize=16cm \epsfbox{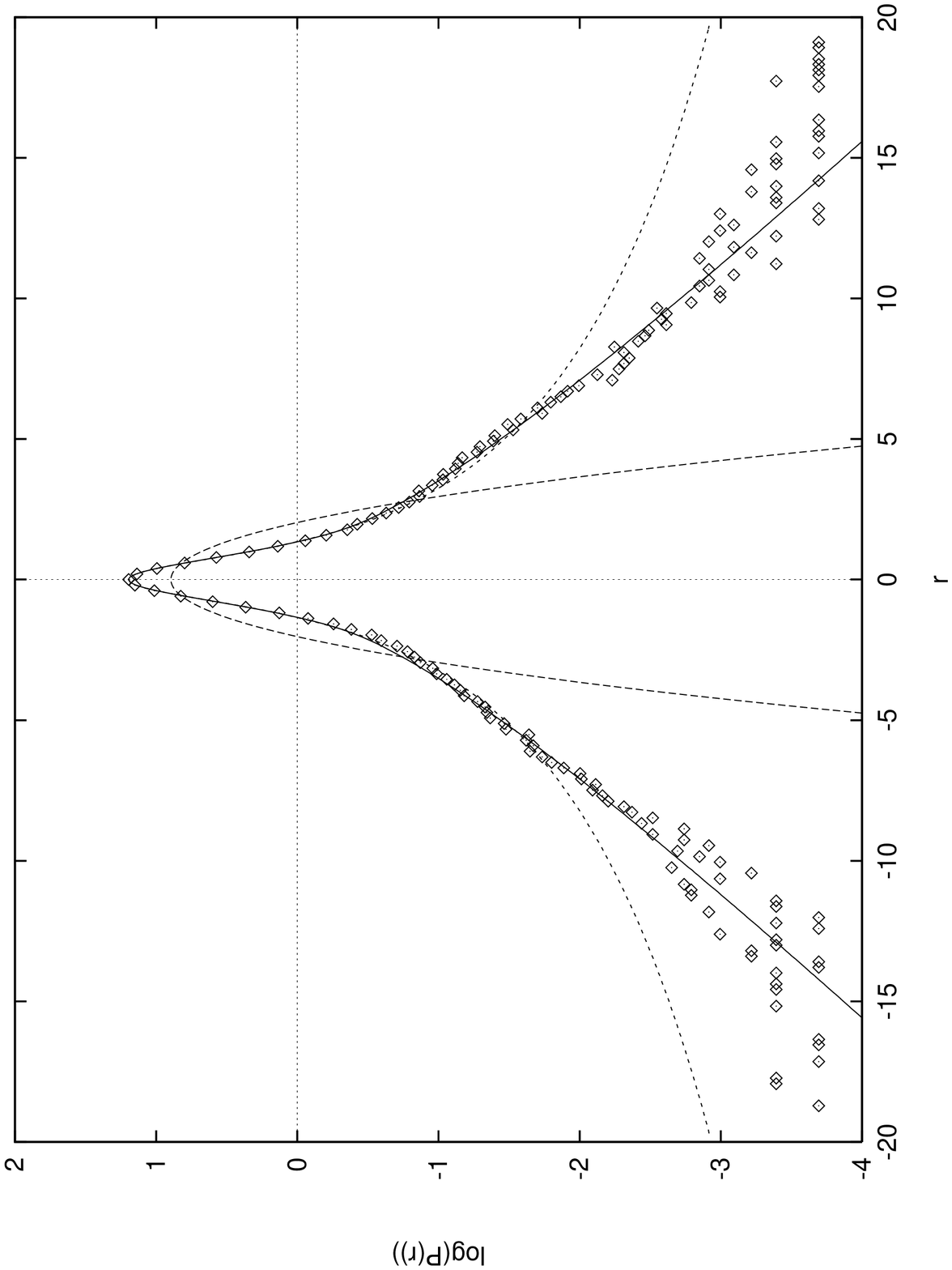}}
\end{figure}


\begin{thebibliography}{99}

\bibitem{Cuthbertson} K.  Cuthbertson, {\it Quantitative financial
economics}, Jonh Wiley $\&$ Sons, 1996;

\bibitem{Elton} E.J.  Elton, M.J.  Gruber, {\it Modern portfolio theory
and investment analysis}, Jonh Wiley $\&$ Sons, 1995;

\bibitem{Hull} J.C.  Hull, {\it Options, futures and other derivatives},
Prentice Hall International, Inc, 1997;

\bibitem{DeLong} J.B.  De Long, A.  Shleifer, L.H.  Summers and R.J.
Waldmann:  Noise Trader Risk in Financial Markets, {\it Journal of
Political Economy}, {\bf 98}, N4, (1990), 703-738;

\bibitem{Bak} P.  Bak, M.  Paczuski, M.  Shubik:  Price variations in a
stock market with many agents, {\it PHYSICA} {A 246}, N.3-4, (1997)
430-453;

\bibitem{Ilinski} K.  Ilinski:  Physics of Finance, to appear in Edited
Volume on Econophysics, Kluwer publishing; available at
http://xxx.lanl.gov/abs/hep-th/9710148;

\bibitem{Dunbar} N.  Dumbar:  Market forces, {\it New Scientist}, N2128,
(1998), 42-45;

\bibitem{Breymann} S.  Ghashghaie, W.  Breymann, J.  Peinke, P.  Talkner
and Y.  Dodge:  Turbulent cascades in foreign-exchange markets, {\it
Nature} {\bf 381}, (1996) 767-770;

\bibitem{Stanley} R.N.  Mantegna and H.E.  Stanley:  Scaling behavior in
the dynamics of an economical index, {\it Nature}, {\bf 376}, (1995)
46-49;

\bibitem{Bouchaud} J.P.  Bouchaud, D.  Sornette:  The Black-Scholes
option pricing in mathematical finance - generalization and extensions
for a large class of stochastic processes, {\it Journal de Physique I
(France)}, {\bf 4} (1994) 863-881;

\bibitem{Australia} A.  Matacz:  Financial Modelling and Option Theory
with the Truncated Levy Process, preprint cond-mat/9710197; available at
http://xxx.lanl.gov/abs/cond-mat/9710197;

\bibitem{Sofianos} G.  Sofianos:  Index Arbitrage Profitability, NYSE
working paper 90-04; {The Journal of Derivatives}, {\bf 1}, N1 (1993);

\bibitem{Peters} E.E.  Peters, {\it Fractional Market Analysis}, John
Wiley \& Sons,Inc., 1995;

\bibitem{Engle} R.F.  Engle: Autoregressive Conditional Heteroscedasticity 
with Estimates of the Variance of United Kingdom Inflation,
{\it Econometrica} {\bf 50}, (1982)
987-1007;

\bibitem{Bollerslev} T.  Bollerslev, R.Y.  Chous, K.F.  Kroner:
ARCH Modeling in Finance: A Review of the Theory and Empirical Evidence, 
{\it J.Econometrics} {\bf 52}, (1992) 5-59.

\end{thebibliography}
\end{document}